    \documentstyle[nato]{crckapb} 
\input epsf.sty

\begin{opening}
\title{CAN MIRA VARIABLES TELL US THE CHEMICAL \protect\\
ABUNDANCES IN STELLAR SYSTEMS?}

\author{Michael Feast}
\institute{Astronomy Department, University of Cape Town \\
(mwf@artemisia.ast.uct.ac.za)}
\author{Patricia Whitelock} 
\institute{South African Astronomical Observatory \\
(paw@saao.ac.za)}
\end{opening}
\runningtitle{ABUNDANCE DISTRIBUTIONS FROM MIRAS}
\begin{document}


\begin{abstract}
 A period-metallicity relation for oxygen-rich Miras derived from globular
clusters is applied to similar variables in the galactic Bulge. The
metallicity distribution in the Bulge thus obtained is in good agreement with
that derived from K~giants. This suggests that the periods of Miras are
useful indicators of metallicity distributions, at least in old systems such
as galactic bulges. For systems such as the LMC the situation is not yet
clear.

\end{abstract}

\section{Introduction}
 Whilst detailed, element by element, abundance analyses can now be extended
to relatively faint stars, the overall metallicity of a star remains an
important characteristic. Even to determine this directly for very faint
stars remains a formidable problem. This is especially so when we wish to
measure the spread of metallicities within a given system, since this
evidently requires data on a large number of stars. In the present paper we
suggest that the periods of Mira variables when suitably calibrated can
provide this type of abundance data for old stellar systems such as the
Bulge region of our own Galaxy. This is of particular interest since the
period determination of Mira variables in the bulges of nearby galaxies, as
well as in highly obscured regions of the Bulge of our own Galaxy, is now
within reach.

\section{The Mira Period-Metallicity Relation}
 It has long been known that Miras (in the present paper we are only
concerned with oxygen-rich Miras) are highly evolved, old, low mass stars
and that their kinematics are a function of period (e.g. Feast 1963). This
indicates that the period of a Mira in the general field is a function of
metallicity or initial mass (or both). It is not possible to address this
problem further in the general field since the complexity of Mira spectra
has so far precluded direct spectroscopic measurements of metallicity for
them. However, it is known (Feast 1981) that for those Miras which are
members of globular clusters there is a relation between the Mira period and
the metallicity of the cluster. This relation was quantified (Feast 1992) to
${\rm [Fe/H]} \propto 3.6 \log P$ (where $P$ is the period). 

\begin{figure}
\epsfxsize=12cm 
\hspace{1mm}\epsfbox{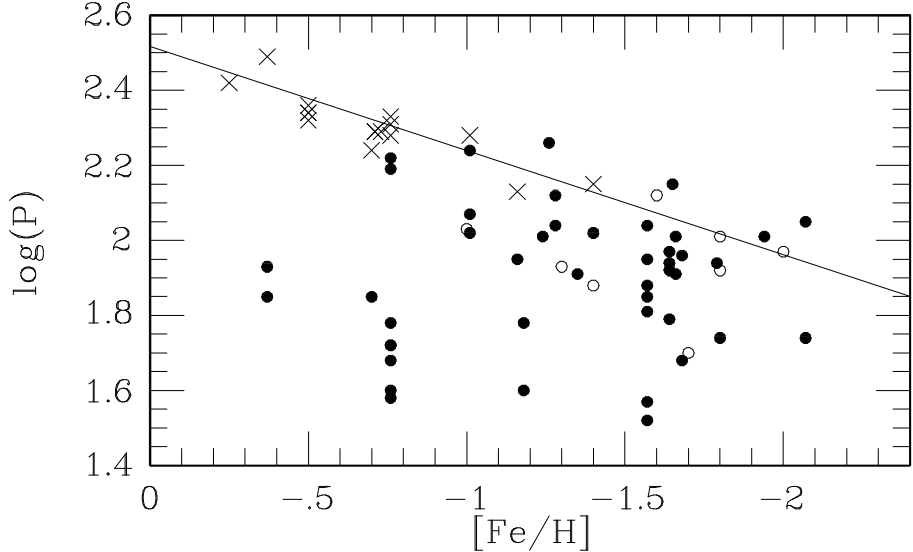} 
  \caption[h]{The pulsation period (in days) of variable stars is shown as a
function of their [Fe/H]. Cluster Miras and SRs are shown as crosses and
filled circles, respectively. Open circles represent field SRs with 
spectroscopically determined [Fe/H] values. }
\end{figure}

The presently available data are shown in fig~1. Here the log of the period
is plotted against metallicity, conventionally denoted as [Fe/H], for Miras
and semiregular (SR) variables in globular clusters.  The data used are
primarily from Frogel and Whitelock (1998), details kindly made available by
Dr J. Frogel. Some additional SR variables in globular clusters are also
included in the plot as well as two additional cluster Miras, V16 in
NGC\,362 (Lloyd Evans 1983) and V1 in NGC\,121 (an SMC cluster). The open
circles represent SR variables in the field of the galactic halo for which
there are spectroscopic determinations of [Fe/H] available (Preston and
Wallerstein 1963, Luck and Bond 1985, Leep and Wallerstein 1981, Giridhar
{\it et al.}, 1998, 1999), taking means where necessary. In a given globular
cluster the SRs lie along an evolutionary track (or more precisely an
isochrone), increasing period going with increasing luminosity (Whitelock
1986). For metal-rich clusters the AGB track terminates with the Miras (if
any) in the cluster. Evidently an evolutionary track in fig~1 is vertical.
The sloping line shown has been drawn by eye to characterize the longest
period at each metallicity and hence may be taken as indicating the
end-point of AGB evolution at each metallicity. The slope of the line was as
determined in the earlier work on Miras (Feast 1992). Various linear fits to
the Mira data are possible depending on how the observations are weighted
and the adopted relative errors in the two co-ordinates. The line shown
however, seems satisfactory for the present purpose as a fit to the Miras,
and has the advantage of giving some weight to the upper limit for the low
metallicity SR variables. In the lower metallicity clusters there are no
Mira variables as normally defined, but the SR variables near the sloping
line are presumably in an equivalent phase of stellar evolution to the Miras
in the more metal-rich clusters. The line is given by the relation:
 \begin{equation}
{\rm [Fe/H]} = 3.60 \log P - 9.06.
\end{equation}
Figure~1 thus indicates that, at least in globular clusters, the
period of a Mira is a good indicator of metallicity.

Frogel and Whitelock (1998) have shown recently that, within the inevitable
uncertainty of small number statistics, the Miras in globular clusters have
another useful property. This is that the ratio of the number of Miras in a
cluster to the total luminosity of the cluster in the infrared (at $K$,
$2.2\mu$m) is independent of metallicity (over the range of metallicities
for which cluster Miras are found).  These results suggest that if the
cluster ${\rm [Fe/H]} - \log P$ relation is applicable to a stellar system,
not only will it yield the range of metallicities in the system, but also,
the relative numbers of Miras of different periods will provide a useful
estimate of the frequency distribution of these metallicities in the parent
population.

\section{Application to the Galactic Bulge}
 We can now test and discuss the cluster ${\rm [Fe/H]}  - \log P$ relation
by applying it to Miras in the galactic Bulge. To do this we have combined
data on Miras in the two Baade windows, Sgr\,I and the NGC\,6522 field. 
Combining the data from the two fields improves the statistics considerably.

\begin{figure}
\epsfysize=7cm 
\hspace{32mm}\epsfbox{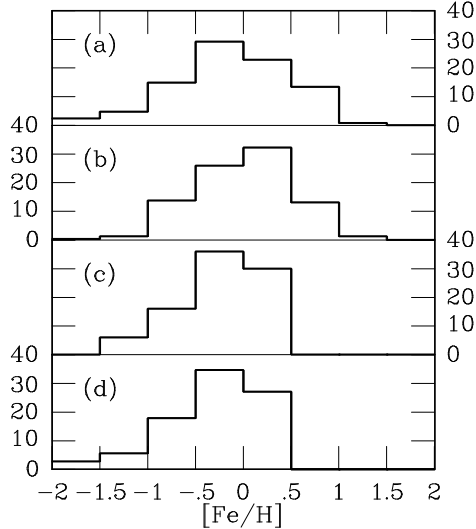} 
  \caption[h]{Histograms of the metallicity distribution of galactic
Bulge stars: (a) Miras (as derived from periods); (b) K~giants (Sadler {\it
et al.}); (c) K~giants (McWilliam and Rich); (d) Miras renormalized as
described in the text. The samples have been normalized to the same
effective total number of objects.}
\end{figure}   

In carrying out an analysis such as this it is essential that the results
are not biased by selection effects. Although the infrared ($K$) and
bolometric luminosities of Miras increase with increasing period, the visual
luminosity decreases because the longer the period, the cooler the star. The
initial South African observations in the Sgr\,I field were on blue
sensitive plates and the variables discovered on them (Oosterhoff and Ponsen
1968) were mainly of short period. The later work was on red sensitive
plates and the proportion of longer period Miras found was larger (Lloyd
Evans 1976). Later still Miras detected by IRAS extended further the long
period coverage (Feast 1986, Glass 1986, Glass {\it et al.}, 1995), see
Glass {\it et al.} fig~2.  For the NGC\,6522 field we have used the periods
from Lloyd Evans. This samples seems essentially complete as it includes all
except one of the IRAS Miras in the field. In the case of the Sgr\,I field
we use the data from Glass {\it et al.} which includes the IRAS Miras
together with periods from Lloyd Evans for Miras not observed by Glass {\it
et al.}

Where necessary, in this first application, we simply extrapolate equation~1
linearly to higher metallicities, outside the range of the globular cluster
Miras. The consequences of this extrapolation will be discussed below.

The frequency distribution of Mira metallicities derived in the above manner
may be compared with the frequency distribution of metallicities of K~giants
in the NGC\,6522 field. For this we have used the values of [Fe/H] derived
by Rich (1988) for 88 stars (his solution~1 rescaled according to equation~6
of McWilliam and Rich (1994)), and the [Fe/H] values for the 262 Bulge K
giants derived from low dispersion spectra by Sadler {\it et al.} (1996) and
plotted in their fig~11. Figure~2 shows this comparison. The Mira distribution
(total number of stars = 112) and the Sadler {\it et al.} K~giant distribution
have been scaled to match the total number of K~giants in the Rich sample.

Figure~2 shows that the distributions of the Miras and the K~giants from
Sadler {\it et al.} agree very well. Sadler {\it et al.} quote a mean [Fe/H]
for their sample of $ -0.11 \pm 0.04$ with a dispersion of $\sigma_{[Fe/H]}
= 0.46$ whilst the Miras give a mean [Fe/H] of $-0.16 \pm 0.06$ and
$\sigma_{[Fe/H]} = 0.59$. These values are not significantly different
for the two samples. The Mira metallicity distribution agrees well with that
of the Rich sample except at the high metallicity end.  The agreement of the
Mira distribution with either the Rich or the Sadler {\it et al.} sample
would be made considerably worse by moving the Miras one box in either
direction.

There are at least three possible (alternative) reasons for the
apparent lack of agreement of the Miras with the Rich sample at the high
metallicity end.

(1) At the high metallicity end, both the Mira and the Sadler {\it et al.}
distributions depend on extrapolations of the (entirely different)
metallicity calibrations used. One can to some extent circumvent possible
problems with the extrapolation by simply comparing the number of Rich K
giants with $\rm [Fe/H] > 0.0$ with the number of Miras in this same
metallicity range. There are 30 Rich K~giants with $\rm [Fe/H] > 0$, whilst
the scaled number of Miras is 37. These numbers are not statistically
different. Thus the results do not preclude the possibility
that the frequency distributions of Rich K~giants and Miras are the same at
all metallicities. A change in the slope of the ${\rm [Fe/H]} - \log P$
relation at high metallicities would suffice to bring the two distributions
into satisfactory detailed agreement.

(2) On the other hand there also seems some possibility that the Rich K
giant distribution might be deficient at the metal-rich end.  This would
occur if all the very metal-rich K~giants were below Rich's magnitude limit. 
In other words, in the magnitude range observed by Rich such metal-rich
stars might be late M giants. In that case a better comparison between the
Rich stars and the Miras is obtained by omitting the long period Miras which
are predicted to have $\rm [Fe/H] > 0.5$, and rescaling. The result of doing
this is shown in fig~2(d). The agreement of the two distributions is
improved in this way.

(3)Finally, it should be borne in mind that the apparent difference between
the Rich K~giant distribution and the other two distributions may be simply
a result of small number statistics.

Our conclusion regarding the Mira metallicity distribution differs somewhat
from that of Frogel and Whitelock (1998). They also obtained a metallicity
distribution from the periods of Bulge Miras (their fig~8). This agrees less
well than ours with the K~giant distribution (either that of Rich or that of
Sadler {\it et al.}) both at high and low metallicities. The main reason for
this is that Frogel and Whitelock used a much steeper ${\rm [Fe/H]} - \log
P$ relation than is consistent with our fig~1. This is due to their relation
being drawn to take into account some metal-rich SR variables which are
below the limiting line that the Miras actually fit.

At first sight our conclusions seem at odds with another of the results
derived by Frogel and Whitelock. They concluded that integrated over all
metallicities, the number of Miras per giant brighter than $\rm M_{bol} =
-1.2$ in the NGC\,6522 field was less than in globular clusters.  However,
at the present time this result must be considered quite uncertain. The
result depends partly on a comparison of the total number of Miras in the
NGC\,6522 field with the number of M type stars brighter than a certain
apparent magnitude in that field and the assumption that stars of the same
apparent magnitude have the same absolute magnitude. Whilst this is true for
a globular cluster it is not so for the galactic Bulge since in the Bulge
the stars are significantly spread along the line of sight.  The spread
($ \sim 2$ mag) in the Mira infrared-PL relation in the Sgr\,I field
(Glass {\it et al.}, 1995) compared with the narrow PL relation in the LMC
(Feast {\it et al.}, 1989) is evidence for such a spread. This will
evidently complicate any analysis. In addition the relative numbers of Miras
to giants in globular clusters is not calculated directly but derived from a
theoretical relation between the total luminosity of a cluster and the
number of giants brighter than a certain value. Whilst our own results are
in satisfactory agreement with the hypothesis that the ratio of numbers of
giants to total luminosity does not depend critically on metallicity, the
absolute calibration of this ratio depends on theoretical giant branch
isochrones.

\section{Is the Mira Period-Metallicity Relation Universal?}
 The evidence just discussed suggests that Mira of a given period in the
galactic Bulge have the same metallicity as Miras of that period in globular
clusters. Is this result universal?

There is evidence of differences between Miras of the same period in
different environments. Glass {\it et al.} (1995) have compared the period -
infrared colour relations for Miras in the Sgr\,I field with those in the
LMC (their fig~4). There is some uncertainty in these relations due to
uncertainties in the correct interstellar absorption to adopt. However, no
adopted relative absorption between the LMC and Bulge fields, with a
standard reddening law, will bring all the period-colour relations into
agreement. If we make the $P -(J-K)$ relations agree then $H-K$ is redder in
Sgr\,I than the LMC at a given period and $J-H$ is bluer. In view of this
result a useful quantity is:
 \begin{equation}
\phi = (J-H)_{0} -(H-K)_{0}.
 \end{equation}
This quantity is rather insensitive to the adopted interstellar reddening.
There is no significant difference in $\phi$ between the Bulge Miras and
those in globular clusters ($\Delta \phi (Glob.Cl. - Bulge) = +0.03 \pm
0.03$). However, a difference, varying somewhat with period, does exist
between the Miras in the LMC and those in the Bulge. Using the stellar
models of Bessell {\it et al.} (1989), Feast (1996) used this difference in
$\phi$ to deduce that the LMC Miras at a given period were metal
deficient compared with those in the Bulge by $\sim 0.4$ dex.  Differences
in $\phi$ reflect differences in the strength of $H_{2}O$ bands at a given
period. This will be affected not only by differences in the atmospheric
[O/H] ratio but also by any difference in the [C/H] ratio (due for instance
to dredge up processes) since oxygen is preferentially locked up in the CO
molecule. Thus whilst it seems safe to assume that the LMC Miras are
deficient in oxygen compared with those in the Bulge the exact amount is
still uncertain. Whether an oxygen deficiency implies a deficiency in other
elements (e.g. iron) is also uncertain.  As summarized by Gilmore and Wyse
(1991) young objects in the LMC seem to have a lower [O/Fe] than similar
galactic objects but it is not known if this extends to older stars. It
remains unclear therefore whether the LMC Miras would fit the
period-metallicity relation of fig~1. However, it is noteworthy that the
Mira in the SMC globular cluster NGC\,121 which is plotted at
$\log P = 2.15$ (Thackeray 1958) and [Fe/H] = --1.4 (Stryker {\it et al.},
1985) in fig~1 fits the adopted relation closely. This suggests that at
least at short periods the Magellanic Cloud Miras fit the adopted relation
\footnote{(1) There is a also a difference in the $(K_{0} - m_{bol}) - \log
P$ relation between the LMC and the Bulge Miras which is probably a
metallicity effect (see Feast and Whitelock 1999).\\ \indent (2) The value
of $\phi$ at a given period is also a function of the pulsation amplitude
(Whitelock {\it et al.} in preparation). This is believed to be due to the
strengthening of the $\rm H_{2}O$ bands as the atmospheric extension is
increased by pulsation. It is unlikely that this is the cause of the LMC -
Bulge difference.}.

The period distribution of O-rich Miras in the LMC peaks at shorter periods
than in the galactic Bulge. Thus the O-rich LMC Miras in Table II of Hughes
and Wood (1989) together with equation~1 above yield a distribution of LMC
O-Mira metallicities peaked between an [Fe/H] of --1.0 and --0.5.  However,
unlike the galactic Bulge there are also carbon Miras in the LMC and it is
not entirely clear how the overall metallicity distribution would be
affected by taking these into account, even if equation~1 applies to the LMC
O-Miras.

\section{Conclusions}
 Our main conclusion is that applying the metallicity scale for Miras as a
function of period set by globular clusters, to Miras in Bulge fields, leads
to a metallicity distribution in good agreement with that shown by Bulge K
giants in the sample of Sadler {\it et al.} The agreement is also good with
the metallicity distribution of the (revised) Rich K~giant sample except at
the very metal-rich (long period) end where a significant extrapolation of
the derived cluster period-metallicity relation is required. A modification
of this extrapolation could remove this discrepancy. Alternatively the K
giant sample of Rich might be deficient in very metal-rich stars.  Thus at
least in the Bulge, the Miras provide a metallicity tracer for a significant
population.


\begin{thebibliography}{}
\bibitem[]{} Bessell, M.S., Brett, J.M., Scholz, M. \& Wood, P.R.
(1989) {\it A\&AS}, {\bf 77}, 1
\bibitem[]{} Feast, M.W. (1963) {\it MNRAS}, {\bf 125}, 367
\bibitem[]{} Feast, M.W. (1981) in: Iben, I. \& Renzini, A., eds.,
{\it Physical Processes in Red Giants}, Reidel, Dordrecht, p.~193
\bibitem[]{} Feast, M.W. (1986) in: Israel, F.P., ed., 
{\it Light on Dark Matter}, Reidel, Dordrecht, p.~339 
\bibitem[]{} Feast, M.W. (1992) in: Bergeron, J., ed., {\it Highlights of
Astronomy}, vol.\ 9, Kluwer, Dordrecht, p.~613
\bibitem[]{} Feast, M.W. (1996) {\it MNRAS}, {\bf 278}, 11
\bibitem[]{} Feast, M.W., Glass, I.S., Whitelock, P.A. \&
Catchpole, R.M. (1989) {\it MNRAS}, {\bf 241}, 375
\bibitem[]{} Feast, M.W. \& Whitelock, P.A. (1999) in: 
Heck, A. \& Caputo, F., ed., {\it Post-Hipparcos Cosmic Candles}, 
Kluwer, Dordrecht, p.~75
\bibitem[]{} Frogel, J.A. \& Whitelock, P.A. (1998) {\it ApJ}, {\bf 116}, 754
\bibitem[]{} Gilmore, G. \& Wyse, R.F.G. (1991) {\it ApJ}, {\bf 367}, L55
\bibitem[]{} Giridhar, S., Lambert, D.L. \& Gonzalez, G. (1998)
{\it PASP}, {\bf 110}, 671
\bibitem[]{} Giridhar, S., Lambert, D.L. \& Gonzalez, G. (1999)
{\it PASP}, {\bf 111}, 1269
\bibitem[]{} Glass, I.S. (1986) {\it MNRAS}, {\bf 221}, 879
\bibitem[]{} Glass, I.S., Whitelock, P.A., Catchpole, R.M.
\& Feast, M.W. (1995) {\it MNRAS}, {\bf 273}, 383
\bibitem[]{} Hughes, S.M.G. \& Wood, P.R. (1989) {\it AJ}, {\bf 99}, 784
\bibitem[]{} Leep, E.M. \& Wallerstein, G. (1981) {\it MNRAS}, {\bf 196}, 543
\bibitem[]{} Lloyd Evans, T. (1976) {\it MNRAS}, {\bf 174}, 169
\bibitem[]{} Lloyd Evans, T. (1983) {\it MNRAS}, {\bf 204}, 961
\bibitem[]{} Luck, R.E. \& Bond, H.E. (1985) {\it ApJ}, {\bf 292}, 559
\bibitem[]{} McWilliam, A. \& Rich, R.M. (1994) {\it ApJS}, {\bf 91}, 749
\bibitem[]{} Oosterhoff, P. Th. \& Ponsen, J. (1968) {\it Bull.\
Ast.\ Inst.\ Neth.\ Sup.}, {\bf 3}, 79 
\bibitem[]{} Preston, G. \& Wallerstein, G. (1963) {\it ApJ}, {\bf 138}, 820
\bibitem[]{} Rich, R.M. (1988) {\it AJ}, {\bf 95}, 828
\bibitem[]{} Sadler, E.M., Rich, R.M. \& Terndrup, D.M. (1996)
{\it AJ}, {\bf 112}, 171
\bibitem[]{} Stryker, L.L., Da Costa, G.S. \& Mould, J.R. (1985)
{\it ApJ}, {\bf 298}, 544
\bibitem[]{} Thackeray, A.D. (1958) {\it MNRAS}, {\bf 118}, 117
\bibitem[]{} Whitelock, P.A. (1986) {\it MNRAS}, {\bf 219}, 525

\end{thebibliography}
\end{document}